\RequirePackage{fix-cm}
\documentclass[smallextended]{svjour3}       
\smartqed  
\usepackage{graphicx}
\usepackage{gensymb}
\usepackage{amsmath}
\usepackage[hyphens]{url}
\usepackage{hyperref}
\usepackage{rotating}
\usepackage{float}
\usepackage[section]{placeins}
\usepackage{amsmath}
\usepackage[version=3]{mhchem}

\graphicspath{{figures/}}

\journalname{Bulletin of Atmospheric Science and Technology}
\begin{document}

\title{Changes in air quality and human mobility in the U.S. during the COVID-19 pandemic}

\titlerunning{Changes in air quality and human mobility in the U.S. during COVID-19}        

\author{Cristina L. Archer$^1$ \and Guido Cervone  \and Maryam Golbazi  \and Nicolas Al Fahel 
\and Carolynne Hultquist 
}


\institute{$^1$Corresponding author: Cristina L. Archer, College of Earth, Ocean, and Environment, University of Delaware, Newark, Delaware 19716, USA, carcher@udel.edu, +1 302 831 6640}

\date{Received: date / Accepted: date}

\maketitle

\begin{abstract}
The first goal of this study is to quantify the magnitude and spatial variability of air quality changes in the U.S. during the COVID-19 pandemic. We focus on two pollutants that are federally regulated, nitrogen dioxide (\ce{NO2}) and fine particulate matter (\ce{PM_{2.5}}). \ce{NO2} is emitted during fuel combustion by all motor vehicles and airplanes. \ce{PM_{2.5}} is emitted by airplanes and, among motor vehicles, mostly by diesel vehicles, such as commercial heavy-duty diesel trucks. Both \ce{PM_{2.5}} and \ce{NO2} are also emitted by conventional power plants, although \ce{PM_{2.5}} almost exclusively by coal power plants. Observed concentrations at all available ground monitoring sites (240 and 480 for \ce{NO2} and \ce{PM_{2.5}}, respectively) were compared between April 2020, the month during which the majority of U.S. states had introduced some measure of social distancing (e.g., business and school closures, shelter-in-place, quarantine), and April of the prior five years, 2015--2019, as the baseline. Large, statistically-significant decreases in \ce{NO2} concentrations were found at more than 65\% of the monitoring sites, with an average drop of 2 parts per billion (ppb) when compared to the mean of the previous five years. The same patterns are confirmed by satellite-derived \ce{NO2} column totals from NASA OMI, which showed an average drop in 2020 by 13\% over the entire country when compared to the mean of the previous five years. \ce{PM_{2.5}} concentrations from the ground monitoring sites, however, were not significantly lower in 2020 than in the past five years and were more likely to be higher than lower in April 2020 when compared to the previous five years. 

The second goal of this study is to explain the different responses of these two pollutants, i.e., \ce{NO2} was significantly reduced but \ce{PM_{2.5}} was nearly unaffected, during the COVID-19 pandemic. The hypothesis put forward is that the shelter-in-place measures affected people's driving patterns most dramatically, thus passenger vehicle \ce{NO2} emissions were reduced. Commercial vehicles (generally diesel) and electricity demand for all purposes remained relatively unchanged, thus \ce{PM_{2.5}} concentrations did not drop significantly. To establish a correlation between the observed \ce{NO2} changes and the extent to which people were actually sheltering in place, thus driving less, we use a mobility index, which was produced and made public by Descartes Labs. This mobility index aggregates cell phone usage at the county level to capture changes in human movement over time. We found a strong correlation between the observed decreases in \ce{NO2} concentrations and decreases in human mobility, with over 4 ppb decreases in the monthly average where mobility was reduced to near zero and around 1 ppb decrease where mobility was reduced to 20\% of normal or less. By contrast, no discernible pattern was detected between mobility and \ce{PM_{2.5}} concentrations changes, suggesting that decreases in personal-vehicle traffic alone may not be effective at reducing \ce{PM_{2.5}} pollution. 

\end{abstract}

\section{Introduction}
\label{intro}






Worldwide, about 91\% of the population is exposed to poor air quality. The World Health Organization (WHO) estimates that on average, 4.2 million people die each year from causes directly attributed to air pollution \cite{whodeath}. Nitrogen dioxide (\ce{NO2}) is one of a group of highly reactive gases known as nitrogen oxides (\ce{NOx}). \ce{NO2} can irritate the human respiratory system and is also harmful to ecosystems by the formation of nitric acid and acid rain \cite{epano2,lin2011}. \ce{PM_{2.5}} is another harmful air pollutant that consists of microscopic particles with a diameter smaller than 2.5 $\mu$m. These particles can pose a great risk to human health because they can penetrate into human lungs and even the bloodstream; \ce{PM_{2.5}} is also often associated with poor visibility \cite{epapm}. \ce{NO2} and \ce{PM_{2.5}} are both primary (i.e., they can be directly emitted into the atmosphere) and secondary (i.e., they can also form after chemical reactions in the atmosphere) pollutants. High concentrations of both are not necessarily found where their emissions are highest, due to processes such as chemical reactions, transport, or diffusion. \ce{NO2} and \ce{PM_{2.5}} are the main focus of this paper because they are among the seven ``criteria" pollutants that are regulated at the federal level by the U.S. Environmental Protection Agency (EPA) via the National Ambient Air Quality Standards (NAAQS). 

The novel coronavirus disease (SARS-CoV-2/COVID-19, COVID-19 hereafter for brevity) was first identified in Wuhan, China, on December 30, 2019 \cite{who,chan2020}. Cases started to spread initially in China but quickly expanded to other countries across the world. COVID-19 was declared a global pandemic in March 2020 \cite{whoEurope}. At the time of this study, over 9 million people have been affected by the virus, with over 470,000 deaths in 213 countries and territories \cite{worldometer,JHU}. COVID-19 first reached the U.S. in February 2020 and since then it has caused over 120,000 deaths in the span of a few months \cite{cdc,JHU}. The death rate of COVID-19 is significantly higher among people with cardiovascular and respiratory illnesses \cite{acc}, which are also strongly connected with air pollution \cite{isaifan2020}. Furthermore, new studies suggest that higher concentrations of air pollutants result in a higher risk of COVID-19 infection \cite{yongjian2020} and mortality \cite{wu2020}.

In the U.S., social distancing measures were implemented state by state with the goal of limiting the spread of the pandemic. In general, closure or non-physical interaction options (e.g., delivery only) were implemented for schools, restaurants, and public places of gathering. Businesses, workers, and types of activities that were deemed essential during the pandemic either continued operating under strict protection measures (e.g., personal protective equipment (PPE), masks) or switched to online work. Non-essential businesses requiring physical presence and interaction closed completely (e.g., hair salons, bars, gyms). The extent of social distancing measures, seriousness of the implementation, and the degree of compliance varied throughout the U.S. Most states announced some level of social distancing orders starting in mid-March, 2020 \cite{bbc}, often including a mandatory quarantine for people diagnosed with or showing symptoms of the coronavirus. By the beginning of April, almost all states had a mandatory shelter-in-place or lockdown order \cite{nystats}. Hereafter, lockdown and shelter-in-place will be used interchangeably. The social distancing measures have led to drastic changes in mobility and energy use and therefore changes in emissions of pollutants. 

Globally, the COVID-19 outbreak is forcing large changes in economic activities \cite{NCAR}. In China, following the strict social distancing measures, transportation decreased noticeably and, as a result, China experienced a drastic decrease in atmospheric pollution, specifically \ce{CO} \cite{NCAR}, \ce{NO2}, and \ce{PM_{2.5}} \cite{manuel2020,NCAR,nasa} concentrations in major urban areas. However, emissions from residential heating and industry remained steady or slightly declined \cite{chen2020}. Using satellite data, Zhang et al. \cite{zhang2020nox} and the National Center for Atmospheric Research \cite{NCAR} reported a 70\% and 50\% decrease in \ce{NOx} concentrations in Eastern China, respectively. Bao and Zhang \cite{bao2020} showed an average of 7.8\% decrease in the Air Quality Index over 44 cities in northern China. Bawens et al. \cite{Bauwensetal2020} and Shi and Brasseur \cite{ShiBra20} reported an increase in \ce{O3} concentrations in the same region. Chen et al. \cite{chen2020}, reported that \ce{NO2} and \ce{PM_{2.5}} concentrations were decreased by 12.9 and 18.9 $\mu$g/m$^3$, respectively. They estimated that this improvement in the air quality of China avoided over 10,000 \ce{NO2}- and \ce{PM_{2.5}}-related deaths during the pandemic, which could potentially outnumber the confirmed deaths related to COVID-19 in China \cite{chen2020}. Other researchers also have proposed that the improvements of air quality during the pandemic might have saved more lives than the coronavirus has taken \cite{dutheil2020,gfeed}. Likewise, Isaifan \cite{isaifan2020} argues that the shutting down of industrial and anthropogenic activities caused by COVID-19 in China may have saved more lives by preventing air pollution than by preventing infection.

European countries, such as France and Italy, experienced a sharp reduction in their air pollution amid COVID-19 \cite{esa}. In Brazil, a significant decrease in \ce{CO} concentrations and, to a lower extent, in \ce{NOx} levels was observed, while ozone levels were higher due to a decrease in \ce{NOx} concentrations in \ce{VOC}-limited locations \cite{dantas2020,nakada2020}. The same findings were observed in Kazakhstan and Spain, respectively \cite{kerimray2020,tobias2020changes}. Likewise, Iran \cite{nemati2020} and India \cite{mahato2020,sharma2020} reported noticeable improvements in air pollution during the pandemic. Le et al. \cite{le2020} looked into the impacts of the forced confinement on \ce{CO2} emissions and concluded that global \ce{CO2} emissions decreased by 17\% by early April compared to the average level in 2019. They believe that the yearly-mean \ce{CO2} emissions would decrease by 7\% if restrictions remain by the end of 2020.

In the United States, as a result of social distancing, states started to experience a dramatic decrease in personal transportation and mobility in general \cite{gao2020}. Personal vehicle transportation decreased by approximately 46\% on average nationwide, while freight movement only decreased by approximately 13\% \cite{inrixtraffic}. Air traffic decreased significantly as well \cite{businessinsider}. On-road vehicle transportation is a main source of \ce{NOx} emissions \cite{nei}. Airports too are usually hot spots for \ce{NO2} pollution \cite{nasa}.

The Houston Advanced Research Center (HARC) \cite{harc} analyzed the daily averages of hourly aggregated concentrations of benzene, toluene, ethylbenzene, and xylenes (BTEX) across six stations in Houston, USA. They reported a decrease in BTEX levels in the atmosphere while an intensified formation on \ce{PM_{2.5}} in the region. Similarly, the New York Times reported huge declines in pollution over major metropolitan areas, including Los Angeles, Seattle, New York, Chicago, and Atlanta using satellite data \cite{NYtimes}.

While a noticeable number of studies have looked into the correlation between lockdown measures amid COVID-19 and air quality in different countries, none has evaluated air quality for the entire United States. The goals of this study are to investigate the magnitude and spatial variability of air quality (\ce{NO2} and \ce{PM_{2.5}}) changes in the U.S. during the COVID-19 pandemic and to understand the relationships between mobility and \ce{NO2} changes. An innovative aspect of this study is that we use an extensive database of ground monitoring stations for \ce{NO2} and \ce{PM_{2.5}} (Section \ref{sec:AQdata}) and a third-party high-resolution mobility dataset derived from cellular device movement (Section \ref{sec:mob}). In addition, we included satellite-retrieved \ce{NO2} information to increase the spatial data coverage (Section \ref{sec:sat}). Whereas most studies rely only on a comparison to 2019, we consider five prior years (2015--2019) to provide a more robust measure.

\section{Data}
\label{sec:data}

\subsection{Air quality data}
\label{sec:AQdata}
 Criteria pollutant concentration data, originally measured and quality-checked by the various state agencies, are centrally collected and made available to the public by the EPA through their online Air Quality System (AQS or AirData) platform \cite{AQS}. For \ce{NO2}, the reported concentrations are one-hour averages, thus 24 records are reported daily (if no records are missing). For \ce{PM_{2.5}}, the reported concentrations are 24-hour averages, thus one value is reported per day. The AQS pre-generated files are updated twice per year: once in June, to capture the complete data for the prior year, and once in December, to capture the data for the summer. The daily files, containing daily-average and daily-maximum of one-hour \ce{NO2} concentrations and 24-hour-average of \ce{PM_{2.5}} concentrations, were downloaded for the years 2015--2019. At the time of this study (May 2020), however, the pre-generated files for April 2020 were not yet available. 

For the year 2020 only, the data source was the U.S. EPA AirNow program \cite{Airnow}, which collects real-time observations of criteria pollutants from over 2,000 monitoring sites operated by more than 120 local, state, tribal, provincial, and federal agencies in the U.S., Canada, and Mexico. As stated on AirNow website, ``these data are not fully verified or validated and should be considered preliminary and subject to change." Of the two types of files available from AirNow, namely AQObsHourly and Hourly, AQObsHourly files were downloaded for March and April 2020 because of their smaller file size (they are updated once per hour, as opposed to multiple times). Texas and New York do not feed \ce{NO2} measurements to Airnow, thus their 2020 \ce{NO2} data were downloaded directly from their state websites \cite{TXNO2,NYNO2}.

The NAAQS for \ce{NO2} and \ce{PM_{2.5}} are based on the comparison of a ``design value", which is a specific statistic of measured concentrations over a specific time interval, against a threshold value as follows:
\begin{itemize}
    \item \ce{NO2}: annual mean of 1-hour concentrations may not exceed 53 parts per billion (ppb);
    \item \ce{NO2}: 98th percentile of 1-hour daily maximum concentrations, averaged over 3 years, may not exceed 100 ppb;
    \item \ce{PM_{2.5}}: annual mean of 24-hour concentrations, averaged over 3 years, may not exceed 12 $\mu$g/m$^3$;
    \item \ce{PM_{2.5}}: 98th percentile of 24-hour concentrations, averaged over 3 years, may not exceed 35 $\mu$g/m$^3$. 
\end{itemize}

Clearly it is not possible to calculate the design values as early as April because neither the annual average nor the 98th percentile can be calculated after only four months. As such, in this study we will use a simple monthly average as the representative metric to compare the concentrations in April 2020 to those in the previous five Aprils. 

\begin{figure}[ht]
    \centering
    \includegraphics[width=\textwidth]{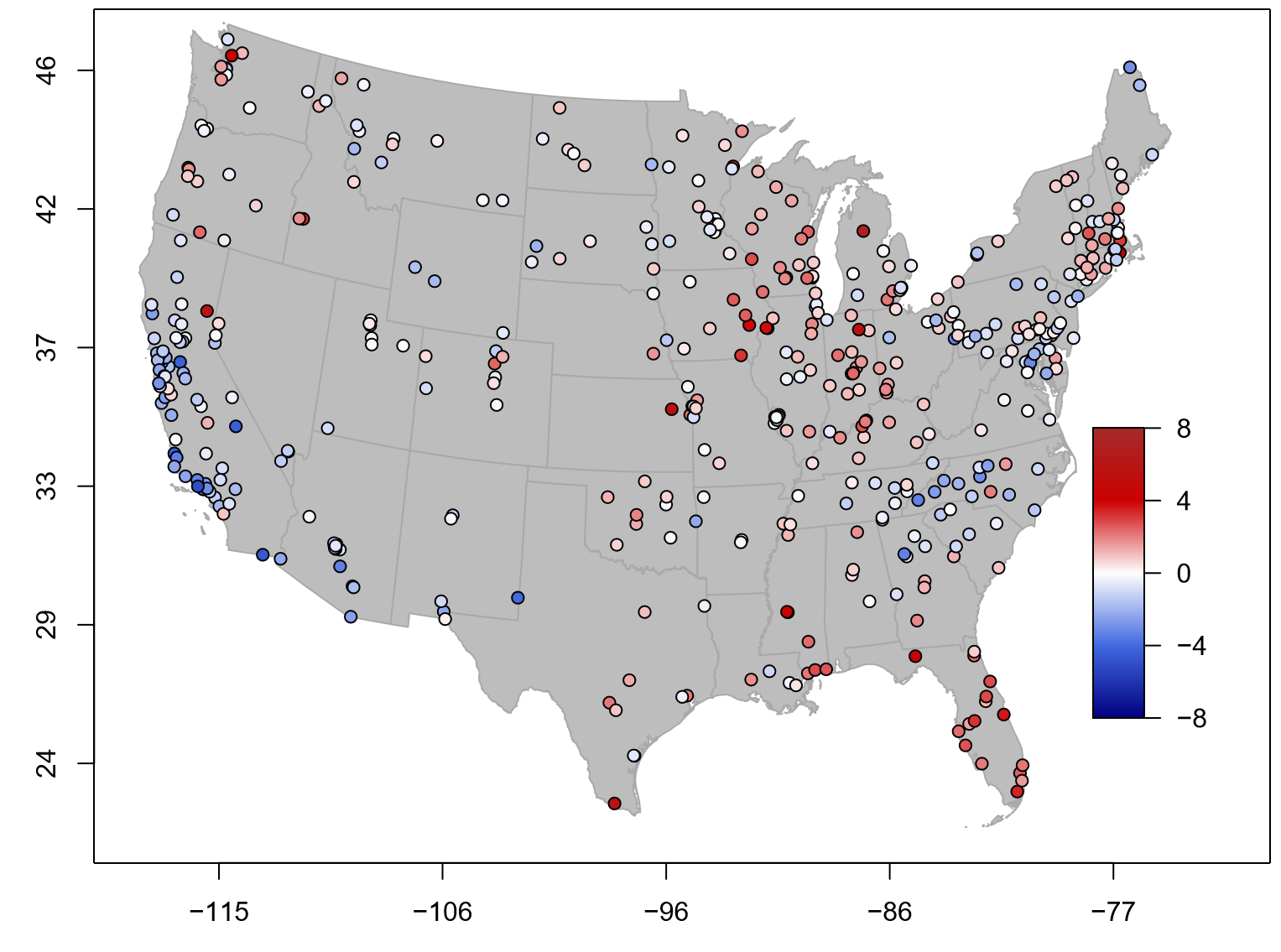}
    \caption{Difference in monthly-average \ce{PM_{2.5}} concentrations ($\mu$g/m$^3$) between April 2020 and the five previous Aprils (2015--2019). Negative values (blue) indicate a decrease in \ce{PM_{2.5}} concentrations in April 2020, vice versa positive values (red) indicate an increase.}
    \label{fig:PM25map}
\end{figure}

An air quality station, whether measuring \ce{NO2} or \ce{PM_{2.5}}, was used in this study only if it reported both in 2020 (through AirNow) and in the five years prior (through AQS). In addition, only air quality stations that were reporting at least 75\% of the time were retained. Note that not all \ce{NO2}-measuring sites also measure \ce{PM_{2.5}}, and vice versa. Of the 426 and 882 sites that measured \ce{NO2} and \ce{PM_{2.5}}, respectively, in April 2020, only 271 and 819 reported at least 75\% of the time, and ultimately only 201 and 480 reported \ce{NO2} and \ce{PM_{2.5}} also in April 2015--2019 for at least 75\% of the time. These are the sites that we will focus on in this study and that are shown in Figures \ref{fig:NO2map} and \ref{fig:PM25map}.

\subsection{Satellite data}
\label{sec:sat}
Satellite observations for \ce{NO2} were acquired using the OMI instrument flying onboard the NASA AURA satellite, and were downloaded using the NASA GIOVANNI portal \cite{giovanni}. Specifically, the Nitrogen Dioxide Product (OMNO2d) was used, which is a Level-3 global gridded product at a 0.25x0.25 spatial resolution provided for all pixels where cloud fraction is less than 30\%. The product comes in two variants, the first measuring the concentration in the total column and the second the concentration only in the troposhere. For this work, the latter measurements were used.

\begin{figure}[ht]
    \centering
    a)\includegraphics[width=8cm]{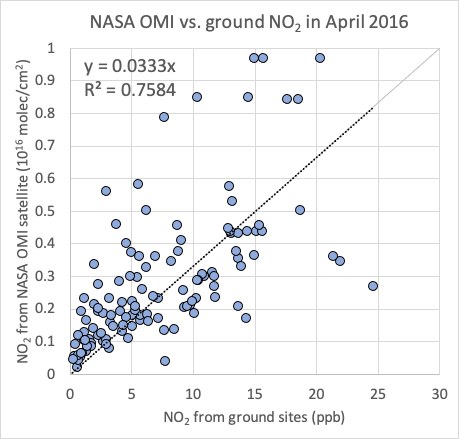}
    b)\includegraphics[width=8cm]{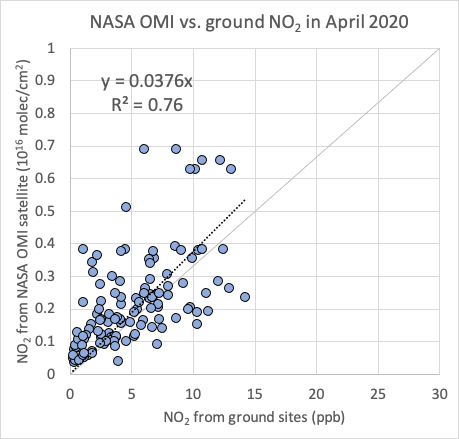}
    \caption{Scatter plots of monthly-average \ce{NO2} concentrations (ppb) from the ground monitoring sites versus monthly-average \ce{NO2} column totals ($10^{16}$ molecules/cm$^2$) retrieved from the NASA OMI satellite at the pixels of the ground monitoring sites during (a) April 2016 and (b) April 2020. }
    \label{fig:sat-stations}
\end{figure}

The satellite-derived \ce{NO2} column totals at the pixels of the ground monitoring sites are well correlated with the \ce{NO2} concentrations recorded at the ground monitoring sites in all years, with R-square values varying between 0.76 and 0.80. As an example, we show the correlation between the two in 2016 and 2020 in Figure \ref{fig:sat-stations}. As such, we can use satellite-derived \ce{NO2} column totals to: 1) confirm the results obtained from the ground monitoring sites and 2) analyze pixels where no ground monitoring sites are available. 

\subsection{Mobility data}
\label{sec:mob}

Mobility measures aim to capture general patterns of observed movement and most available data products today utilize mobile device activity as a proxy. While policy makers set social distancing guidance, there are various policies enacted and various degrees to which policies are followed throughout the country. We seek to observe actual patterns of movement by using a dataset developed by Descartes Labs \cite{descartes} that provides an aggregated mobility measure based on anonymized and/or de-identified mobile device locations. Mobility is essentially a statistical representation of the distance a typical member of a given population moves in a day. Descartes Labs calculated the farthest distance apart recorded by smartphone devices utilizing select apps (with location reporting enabled) for at least 10 uses a day, spread out over at least 8 hours in a day, with a day defined as 00:00 to 23:59 local time \cite{Warren2020}. The maximum distance for each qualifying device is tied to the origin county in which the anonymized user is first active each day. Aggregated results at the county level are produced as a statistical measure of general travel behavior. 

Mobility data are ultimately provided as percent of normal, i.e., the ratio of aggregated mobility during each day of the COVID-19 pandemic over that of the baseline (17 February--7 March 2020). Note that the baseline period is in late winter 2020, whereas the period of focus in this paper is April 2020, in spring. As such, a fraction of the differences in mobility may be due to differences in weather and/or climate rather than to COVID-19 restrictions. We did not attempt to correct for this type of bias. Another caveat, noted by the the producers of the data \cite{Warren2020}, is that the raw data used to calculate mobility are available for only a small fraction of the total number of devices (a few percent at most), thus the resulting mobility may or may not truly represent the average behaviour in each county. Nonetheless, the effects of these sampling errors are expected to be small. The mobility data are made freely available by Descartes Labs at the U.S. county level \cite{gitdata}.

\section{Results}
\label{sec:results}

\subsection{Observed air quality changes}
\label{sec:conc}

In the rest of this paper, we will compare the monthly-average of the pollutant of interest -- \ce{NO2} or \ce{PM_{2.5}} -- during the month of April 2020 to the average of the five monthly-averages during April of the years 2015 through 2019. There are two reasons for this choice. First, using five years to establish a reference is more meaningful than, say, using just the year 2019, because year-to-year variability can occur regardless of the pandemic. In fact we found that, in general, the year 2019 was relatively clean when compared to the previous five, thus a comparison between April 2020 and April 2019 may under-estimate the true impact of COVID-19. Second, although the monthly-average is not the design value for either \ce{NO2} or \ce{PM_{2.5}}, it is a value that is representative of the overall air quality during the entire month of April. Alternative metrics, such as the monthly maximum, are more representative of extreme circumstances, like wildfires, that are not necessarily associated to COVID-19.     

\begin{figure}[ht]
    \centering
    \includegraphics[width=\textwidth]{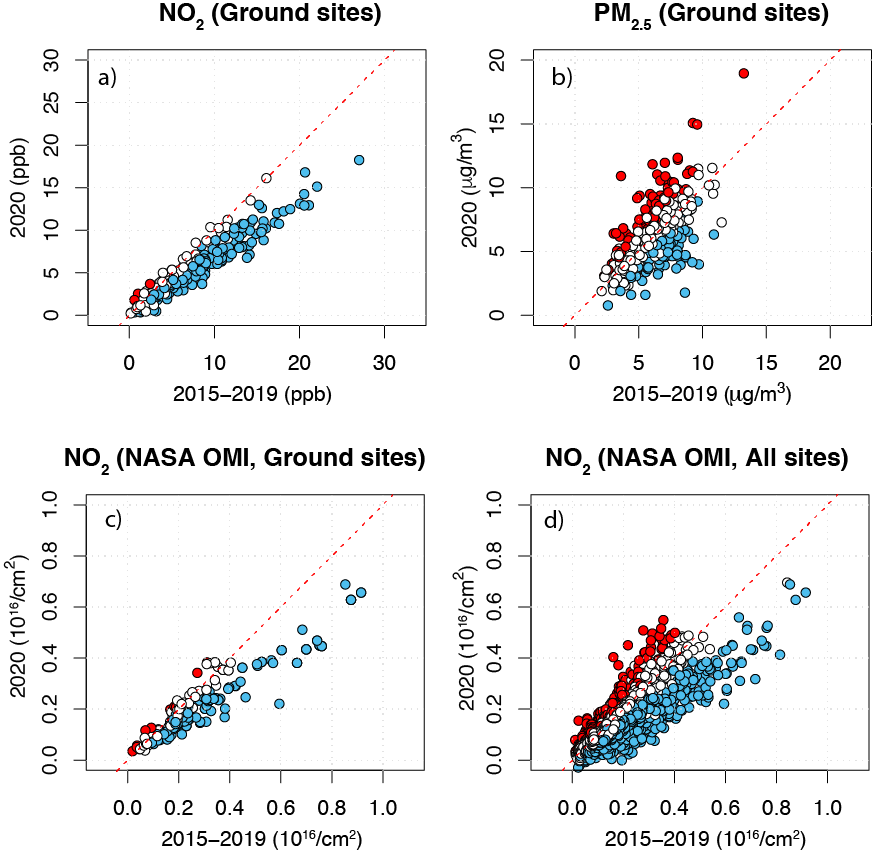}
    \caption{Scatter plots of: (a) \ce{NO2} and (b) \ce{PM_{2.5}} observed concentrations at ground monitoring sites, and \ce{NO2} column totals from NASA OMI satellite at (c) pixels of the ground monitoring sites and (d) all pixels during April of 2020 (y-axis) versus April of the previous five years 2015--2019 (x-axis). Blue-filled markers represent sites for which the values in April 2020 were lower than in any April of 2015--2019; red-filled markers represent sites for which the values in April 2020 were higher than in any April of 2015--2019. }
    \label{fig:scatter}
\end{figure}

Starting with \ce{NO2}, the April 2020 averages were generally below the April 2015--2019 average at the ground monitoring sites, as most sites lay below the 1:1 line in Figure \ref{fig:scatter}b. In addition, 65\% of the sites were characterized by \ce{NO2} concentrations in 2020 that were lower than those in all of the previous five years (for the month of April). Only a few sites (5 in total, $<2\%$) experienced \ce{NO2} concentrations in 2020 that were higher than those in all of the previous five years (for the month of April). The average drop in \ce{NO2} concentrations was -2.02 ppb (Tables \ref{tab:NO2stats} and \ref{tab:stats}).

The same pattern is confirmed in the satellite-derived data. Out of the 227 pixels with ground monitoring sites, a total of 127 (56\%) exhibited lower \ce{NO2}in 2020 than in the previous five years and only 5\% higher (Figure \ref{fig:scatter}b). Once all 14,706 pixels with valid satellite retrievals all over the country are considered, a similar pattern of lower \ce{NO2} column totals in 2020 than in the five previous years emerges from these data too (Figure \ref{fig:scatter}c), but with 28\% of the pixels lower in 2020 than in the previous five years and 5\% higher (for the month of April, Table \ref{tab:stats}).    

\begin{figure}[ht]
    \centering
    \includegraphics[width=\textwidth]{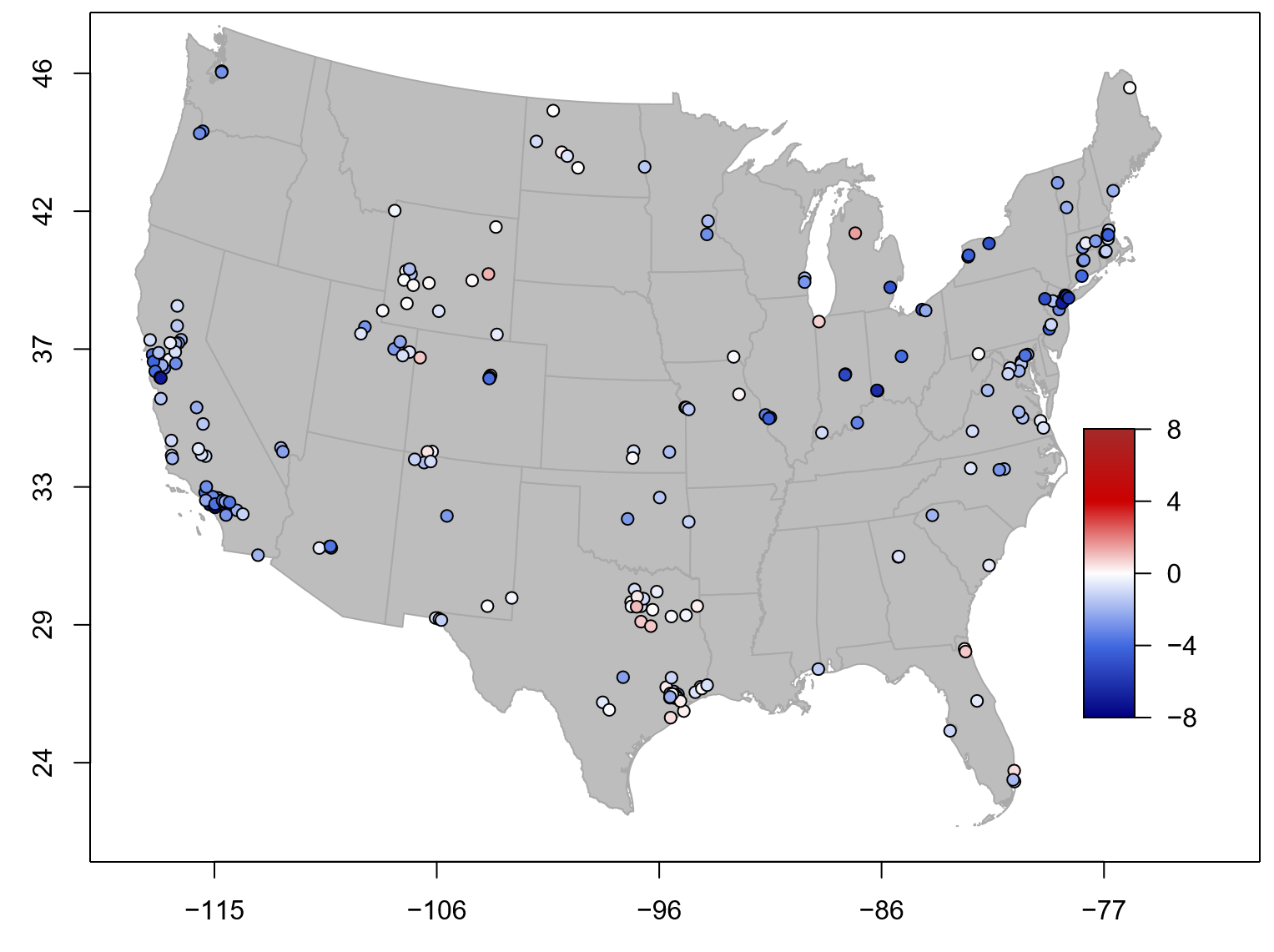}
    \caption{Difference in monthly-average \ce{NO2} concentrations (ppb) between April 2020 and the five previous Aprils (2015--2019). Negative values (blue) indicate a decrease in \ce{NO2} concentrations in April 2020, vice versa positive values (red) indicate an increase.}
    \label{fig:NO2map}
\end{figure}

In terms of spatial variability, Figure \ref{fig:NO2map} shows that, although \ce{NO2} reductions were recorded all over the country, the highest decreases were observed in California and the Northeast, where the shelter-in-place measures started earlier (March 11 for California, the earliest in the country, and March 22 for New York, third earliest \cite{nystats}) and lasted longer (both states still have major restrictions in place as of June 10, 2020 \cite{WaPost}). Noticeable exceptions were North Dakota and Wyoming, where either no significant decreases or actual small increases in \ce{NO2} concentrations were observed. North Dakota enforced no shelter-in-place measures and in Wyoming only the city of Jackson implemented a stay-at-home order as of April 20, 2020 \cite{nystats}. However, as discussed in Section \ref{sec:mob}, actual people's mobility, as opposed to state ordinances, is a better metric to understand the real effect of COVID-19 on air quality because not everybody in all counties followed the state- or county-restrictions all the time.   

\begin{figure}[ht]
    \centering
    \includegraphics[width=\textwidth]{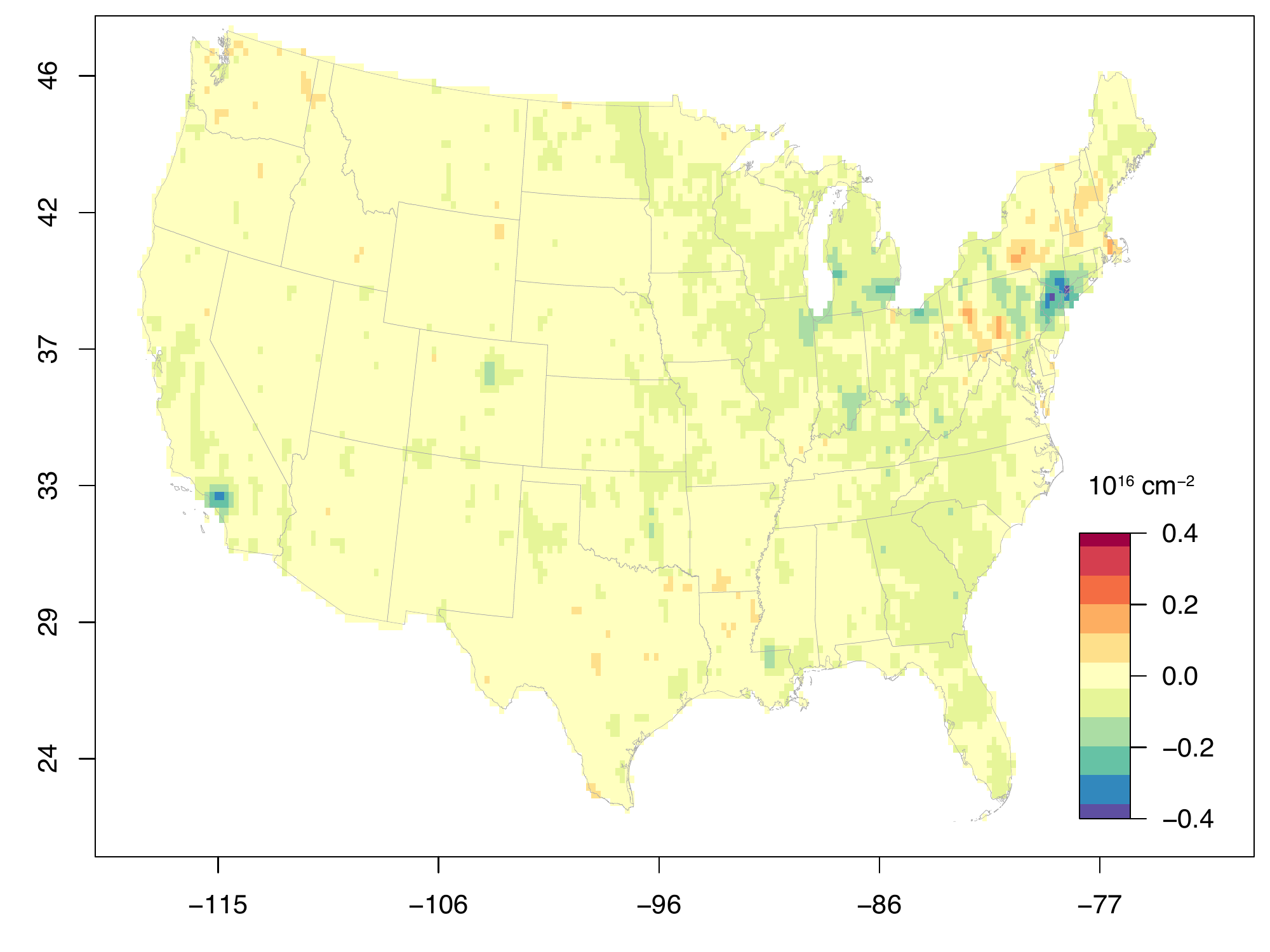}
    \caption{Difference between the NASA OMI \ce{NO2} monthly-average column totals ($10^{16}$ molecules/cm$^2$) in 2020 and in 2015--2019 for the month of April. The ``hotspot" of reduced \ce{NO2} in the Northeast is apparent. }
    \label{fig:NO2sat}
\end{figure}

Figure \ref{fig:NO2map} was useful because it included actual \ce{NO2} concentrations measured near the ground. However, the spatial coverage was sparse and urban areas were over-sampled compared to rural areas. This weakness is addressed via the NASA OMI satellite data, which are shown in Figure \ref{fig:NO2sat} as the difference between the monthly average of \ce{NO2} column total in 2020 and that in 2015--2019 for the month of April. The regions with low coverage of ground concentration of \ce{NO2} and mobility in the Midwest are generally characterized by near-normal \ce{NO2} column totals. The Northeast hotspot of low mobility is also a hotspot of low \ce{NO2}, consistent with \cite{Bauwensetal2020}, although it is surrounded by patches of above-normal values that were not detectable from the ground monitoring stations. The Los Angeles area is another hotspot of \ce{NO2} decreases, as for low mobility.        
For \ce{PM_{2.5}}, the ground monitoring stations depict a completely different response to COVID-19. Whereas most \ce{NO2} sites were laying below the 1:1 line (Figure \ref{fig:scatter}a), the majority of \ce{PM_{2.5}} sites laid above it (Figure \ref{fig:scatter}b), indicating an overall increase in monthly-average \ce{PM_{2.5}} in the country in April 20202 with respect to the previous five years. Only 18\% of the sites reported concentrations of \ce{PM_{2.5}} that were lower in 2020 than in the previous five years (in the month of April), while 24\% of the stations reported the highest levels in 2020 compared to the previous five years (for the month of April). The average increase in \ce{PM_{2.5}} concentrations was small, +0.05 $\mu$g/m$^3$ (Tables \ref{tab:PM25stats} and \ref{tab:stats}).

In summary, we report a large decrease (-2.02 ppb, or 27\%) in monthly-average \ce{NO2} concentrations across the U.S. ground monitoring stations, confirmed by the satellite-derived \ce{NO2} column total decrease of 7.1 $\times 10^{14}$ molecules/cm$^2$ (or 24\%) at the pixels of the ground monitoring stations during April of 2020 when compared to April of the previous five years. When all the pixels with valid data were included, a drop of
2.4$\times 10^{14}$ molecules/cm$^2$ (or 13\%) during April of 2020 was observed when compared to April of the previous five years (Table \ref{tab:stats}). The monthly-average of \ce{PM_{2.5}}, however, increased slightly on average (+0.05 $\mu$g/m$^3$ when compared to the previous five-year average) during the same period (Table \ref{tab:stats}). In the next Section \ref{sec:stat}, we try to explain the reasons for these differences.

\subsection{Observed mobility changes}
\label{sec:mobchanges}

\begin{figure}[ht]
    \centering
    a)\includegraphics[width=8cm]{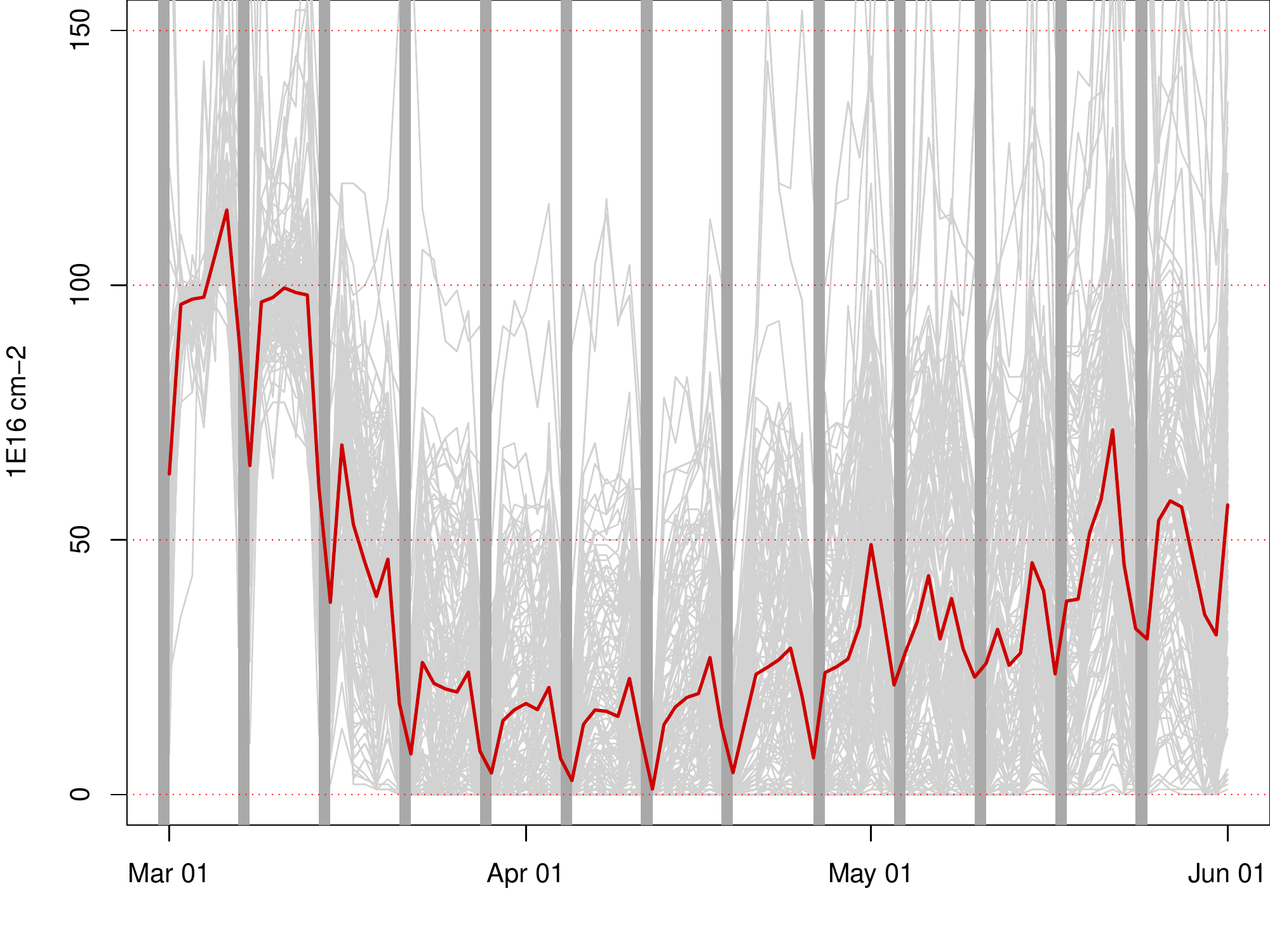}
    b)\includegraphics[width=8cm]{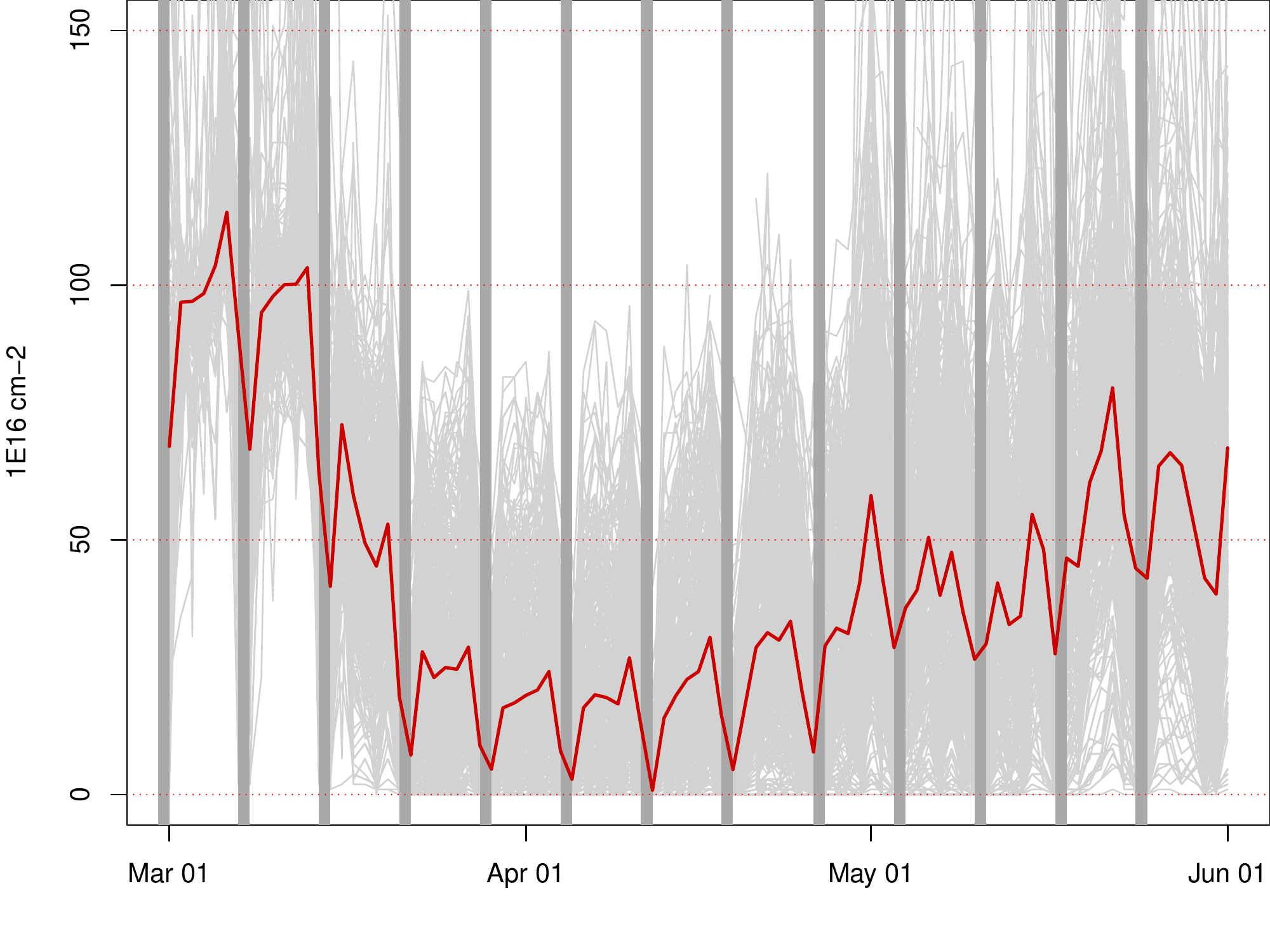}
    \caption{Mobility, expressed as percent of normal, at the locations of the ground stations monitoring (a) \ce{NO2} and (b) \ce{PM_{2.5}} during March--May 2020. The average of all the stations is shown in red.}
    \label{fig:spaghetti}
\end{figure}

Time series of mobility data at the counties with \ce{NO2} ground monitoring sites are shown in Figure \ref{fig:spaghetti}a and at the counties with \ce{PM_{2.5}} ground monitoring sites in Figure \ref{fig:spaghetti}b. Only a few counties had both types of monitoring sites, thus the counties included in the two figures are generally different. Yet, the patterns are very similar. First of all, mobility on average dramatically dropped starting in the second half of March, reaching values around 20\%  by April, and then started to recover in May, as some states reopened for business or relaxed the shelter-in-place measures \cite{WaPost}. Second, a distinct minimum in mobility during the month of April is clearly visible, which confirms that this month was the most relevant for air quality impacts from COVID-19. There is some variability around this general behaviour, but nonetheless only a few counties barely reached normal mobility in April. Lastly, the typical traffic reduction during the weekends is confirmed in the mobility data, regardless of the pandemic. This adds confidence to the use of mobility data as proxies for people's actual behaviours. 

\begin{figure}[ht]
    \centering
    a)\includegraphics[width=\textwidth]{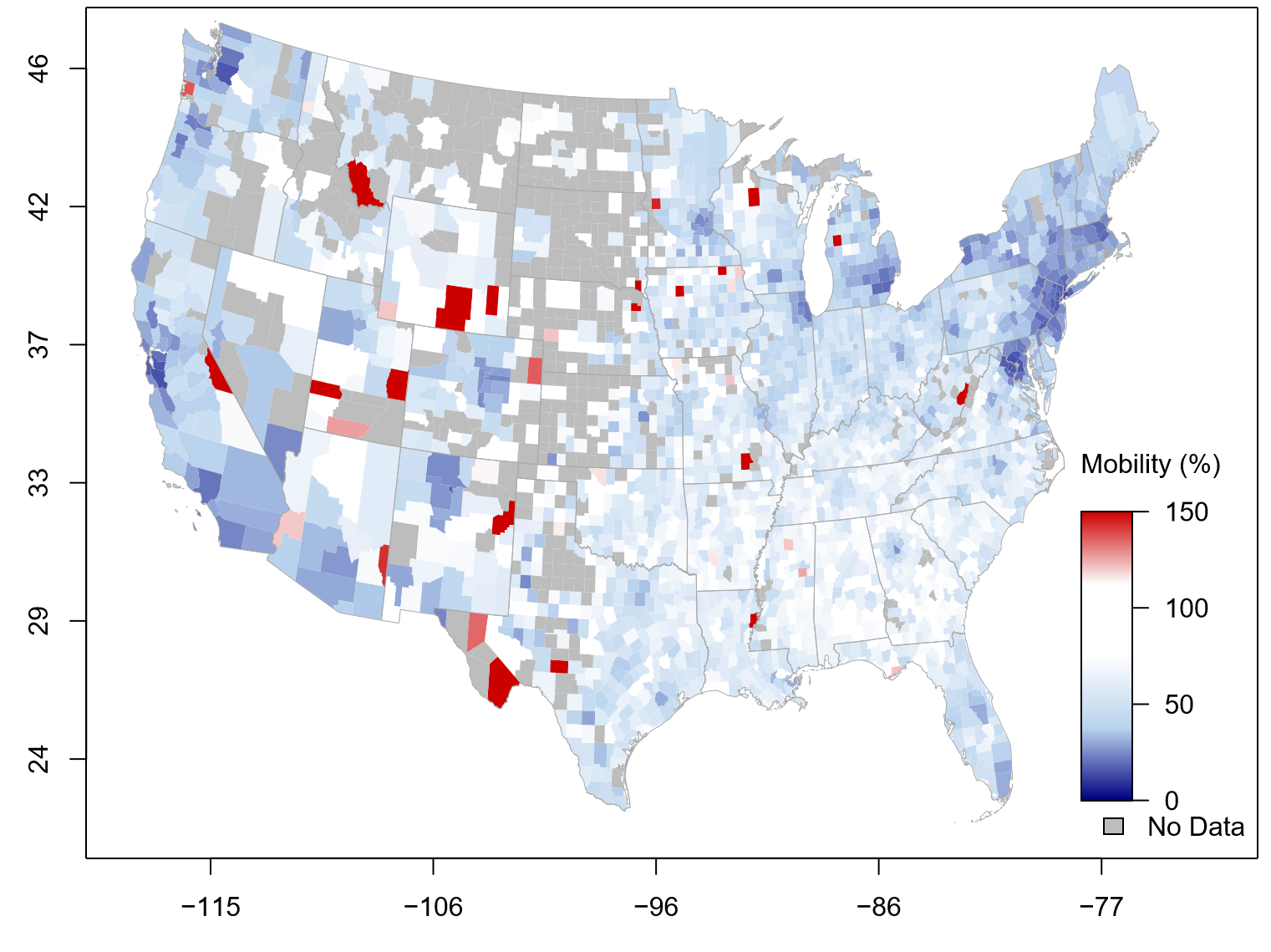}
    b)\includegraphics[width=\textwidth]{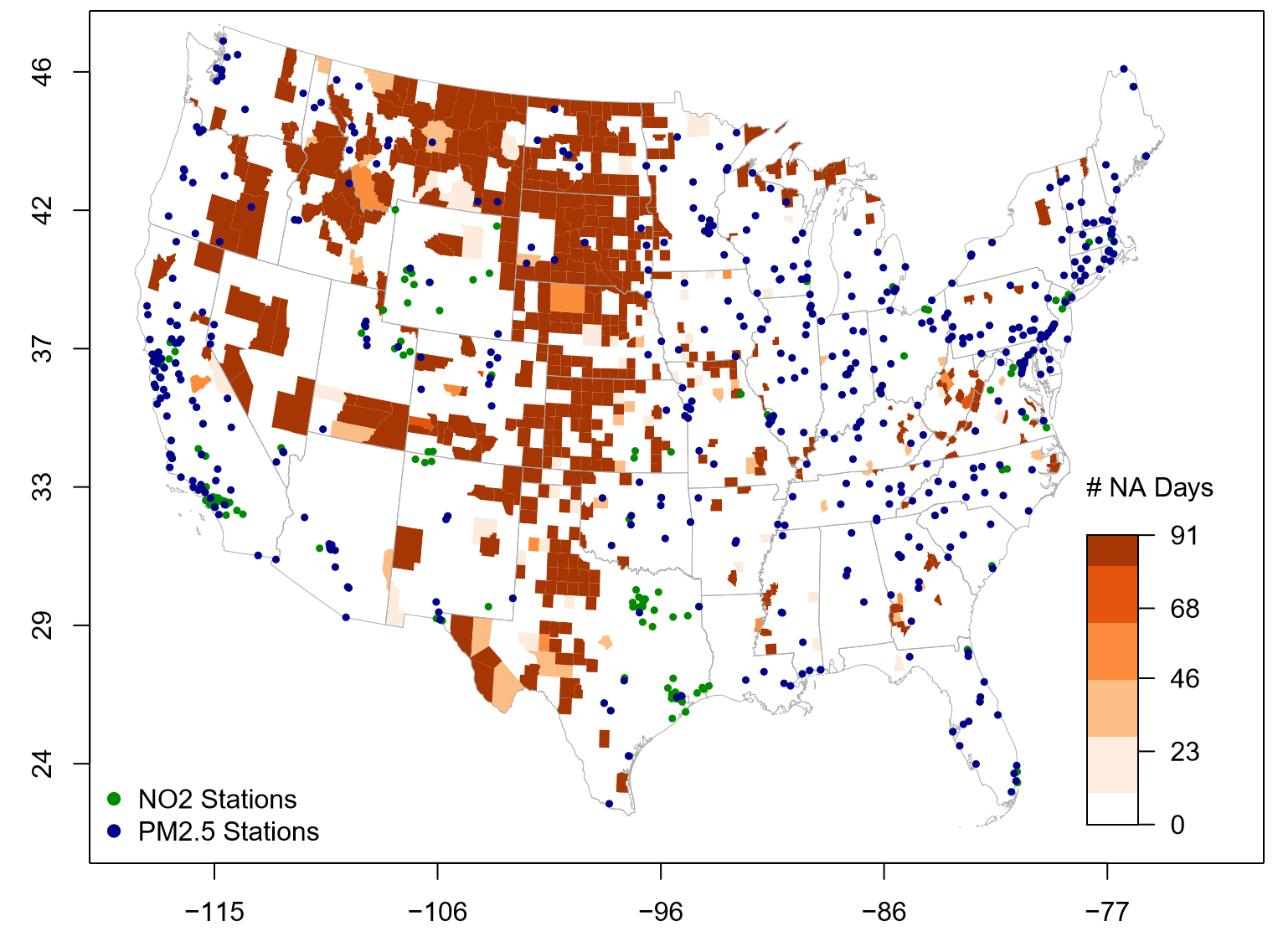}
    \caption{Spatial distribution of (a) mobility, expressed as percent of normal, in April 2020 and (b) mobility data availability, expressed as number of missing days, during March--May 2020.}
    \label{fig:mobility_avg}
\end{figure}

In terms of spatial variability, changes in mobility during COVID-19 in the U.S. were not uniform, although in general mobility was reduced in most states (Figure \ref{fig:mobility_avg}a). Note the high count of non-valid data in many counties in the Midwest (Figure \ref{fig:mobility_avg}b), possibly due to low population density. However, the ground monitoring stations of both \ce{NO2} and \ce{PM_{2.5}} are generally located in counties with high data availability. In general, the strongest decreases in mobility are found around large urban areas throughout the country, e.g., the Northeast corridor from Washington D.C. to Boston; the San Francisco and Los Angeles areas in California; Seattle in the Northwest; and Chicago. A few isolated counties experienced increases in mobility (in red in Figure \ref{fig:mobility_avg}a). Wyoming stands out as one of the few states with no significant decreases in mobility, consistent with the lack of shelter-in-place measures \cite{nystats}.     

\subsection{Relationships between air quality and mobility changes}
\label{sec:stat}

To better interpret the relationship between mobility and the air pollutant of interest, either \ce{NO2} or \ce{PM_{2.5}}, the mobility data were divided into bins, based on the monthly-average (in April 2020) of the mobility in the county where each ground monitoring site was located. For most cases, there was only one ground monitoring site per county. But in some cases, such as Los Angeles county in California for \ce{NO2} or Maricopa county in Arizona for \ce{PM_{2.5}}, multiple monitoring sites were located in the same county and therefore they were all paired to the same mobility value. The change in monthly-average concentration of the pollutant between April 2020 and the five previous Aprils was then calculated for each mobility bin. 

\begin{figure}[ht]
    \centering
    a)\includegraphics[width=12cm]{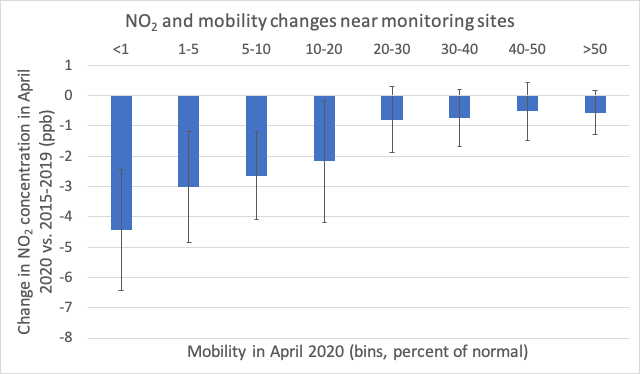}
    b)\includegraphics[width=12cm]{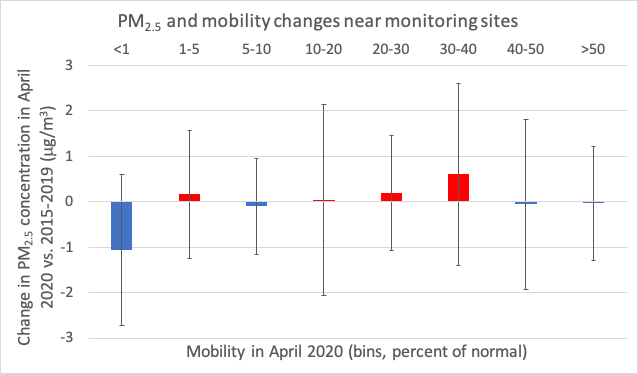}
    \caption{Changes in monthly-average concentrations of (a) \ce{NO2} and (b) \ce{PM_{2.5}} near  ground monitoring sites during April of 2020 versus April of the previous five years as a function of mobility index bins in April 2020. }
    \label{fig:bins}
\end{figure}

Starting with \ce{NO2}, there is a clear relationship with mobility (Figure \ref{fig:bins}a). Large and negative changes in \ce{NO2} concentrations, of the order of -4 ppb, were found at locations where mobility was basically halted, i.e., where it was less than 1\% of normal in April 2020, as in full lock down. As mobility increased, the \ce{NO2} benefits decreased, although not linearly. For example, decreases by 2--3 ppb in \ce{NO2} concentrations occurred where mobility was restricted but not to a full lock down (i.e., between 1\% and 20\% of normal). Past 20\%, the changes in \ce{NO2} concentrations were still negative and significant, but not large, less than 1 ppb on average. This suggests that \ce{NO2} responds modestly to changes in mobility that are not large, but then, if mobility is reduced dramatically (i.e., by at least 80\%, thus it is down to 20\% of normal), large decreases in \ce{NO2} can occur. 

With respect to \ce{PM_{2.5}}, there is no obvious relationship between the reductions in mobility and the observed concentrations (Figure \ref{fig:bins}b). Only for the most extreme mobility reductions, i.e., the bin with $<$1\% mobility, which indicates that the entire population was sheltering at home for the entire month of April, \ce{PM_{2.5}} concentrations decreased by about 1 $\mu$g/m$^3$. After the first bin, as mobility increased, both increasing and decreasing concentrations of \ce{PM_{2.5}} were found, with large standard deviations and no discernible pattern. We conclude that the changes in \ce{PM_{2.5}} were not directly caused by changes in people's mobility. 

How can we reconcile the clear relationship of \ce{NO2} with mobility with the lack thereof for \ce{PM_{2.5}}? The hypothesis we put forward is that the shelter-in-place measures affected mostly people's driving patterns, thus passenger vehicle -- mostly fueled by gasoline -- emissions were reduced and so were the resulting concentrations of \ce{NO2}. Commercial vehicles (generally diesel) and electricity demand for all purposes (often provided by coal-burning power plants), however, remained relatively unchanged, thus \ce{PM_{2.5}} concentrations did not drop significantly and did not correlate with mobility. 
To test this hypothesis, in a subsequent study we will use a photochemical model, coupled with a numerical weather prediction model, which we will run with and without emissions from diesel vehicles, while keeping everything else the same. The difference between the concentrations of the pollutants in the two cases will be attributable to diesel traffic alone. Similarly, we will be able to reduce emissions from other sectors, to reflect the effect of COVID-19 on other aspects of life, such as air traffic or business closures. 

\section{Conclusions and future work}

This study analyzed the effects of COVID-19 on air quality, more specifically \ce{NO2} and fine particulate \ce{PM_{2.5}} concentrations, in the U.S. Although different states introduced different levels of shelter-in-place and social distancing measures at different times, by the beginning of April 2020 all states but a few had adopted some restrictions. As such, the analysis focused on the month of April 2020, which was compared to April of the previous five years, 2015--2019. 

Two types of measurements were used, \ce{NO2} and \ce{PM_{2.5}} concentrations from the ground monitoring stations -- maintained by the states -- and satellite-derived \ce{NO2} column totals in the troposphere. Although the two measurements are not identical, they are strongly correlated with one another because the near-ground concentrations of \ce{NO2} are the dominant contributors to the tropospheric column total. 

To quantify social distancing, we used the mobility index calculated and distributed by Descartes Labs. Their algorithms account for people's maximum distance travelled in a day by tracing the user's location multiple times a day while using selected apps. Mobility is represented as a percent value, such that 100\% means normal conditions, i.e., those during the period of 17 February -- 7 March 2020.  

We found that \ce{NO2} levels decreased significantly in April 2020 when compared to April of the five previous years, by up to 8 ppb in the monthly average at some locations. On average over all U.S. monitoring sites, the decrease in \ce{NO2} levels was between 24\% (from satellite) and 27\% (from ground stations). The decreases in \ce{NO2} were largest where mobility was reduced the most, with a direct, although not linear, relationship between the two. In terms of spatial variability, hotspots of reduced \ce{NO2} concentrations in the Northeast and  California coincided perfectly with hotspots of reduced mobility. Vice versa, states where social distancing measures were minimal experienced the smallest reduction in \ce{NO2}, e.g., Wyoming and North Dakota.    

By contrast, the concentrations of \ce{PM_{2.5}} did not decrease significantly during the same period and even reached unprecedented high values at about a fifth of the sites. In addition, changes in \ce{PM_{2.5}} concentrations were not correlated with changes in people's mobility, neither spatially nor as aggregated statistics. 

We propose that the different response to reduced people's mobility between \ce{NO2} and \ce{PM_{2.5}} could be explained by the fact that commercial vehicles (including delivery trucks, buses, trains), generally diesel fueled, remained more or less in circulation, while passenger vehicles, gasoline fueled, dropped dramatically due to COVID-19. \ce{PM_{2.5}} emissions are much larger from diesel than from gasoline vehicles. In addition, other sources of \ce{PM_{2.5}} emissions, like power plants, did not decrease. We plan to verify this hypothesis in a subsequent study using a photochemical model coupled with a numerical weather prediction model, as described in Section \ref{sec:stat}. 

As far as we know, this is the first study to use ground monitoring stations to assess the effects of COVID-19 on air quality in the U.S. Satellite-derived \ce{NO2} column totals have been used in a few previous studies, but none looked at the correlation between the two types of \ce{NO2} measurements. Another innovation of this study is the use of mobility data, which are an excellent proxy for actual people's behaviour, as opposed to the state or county regulations, which may or may not be fully followed by people.   

This analysis has also numerous limitations. First of all, we paired mobility data and pollutant concentrations at the county level, thus we implicitly assumed that the measured concentration and county-average mobility were representative of the entire county. For large counties, especially those with also low population density, this assumption may not hold. The second implicit assumption of our pairing is that local mobility affects local pollution only and, vice versa, that local pollution is affected by local mobility only. In other words, we are neglecting the effects of transport and chemical reactions, which could cause either an increase or a decrease of pollution regardless of the local mobility change in the county of interest. For example, consider the case that the prevailing wind is such that a county is located downwind of an airport. If the airport was shutdown during the pandemic, that county would see a reduction of \ce{NO2} and \ce{PM_{2.5}} concentrations even if no social distancing measures were in place. Another limitation is that we looked at people's mobility as the only factor explaining \ce{NO2} concentration changes, whereas \ce{NO2} emissions changed also in response to business and school closures, air traffic reductions, among many others sources. Lastly, we focused on two pollutants only, \ce{NO2} and \ce{PM_{2.5}}, because of time constraints; future work will include other regulated compounds, such as ozone and carbon monoxide. 

\begin{acknowledgements}
The authors would like to thank Descartes Labs for providing the mobility data free of charge. The air quality and satellite-derived data were provided by numerous federal and state agencies, listed in the references. 
\end{acknowledgements}

%
\section*{Conflict of interest}
On behalf of all authors, the corresponding author states that there is no conflict of interest.

\bibliographystyle{spmpsci}      

\bibliography{covid19}   

\begin{table}[ht]
\centering
\caption{Monthly-average of \ce{NO2} concentrations (ppb) by state in the month of April of the years 2015--2020.}
\label{tab:NO2stats}
\begin{tabular}{l|r|rrrrr|r}
State&N. Sites&2015&2016&2017&2018&2019&2020\\
\hline
Arizona&4&12.49&11.68&13.58&12.82&10.33&9.06\\
California&56&9.73&9.23&8.79&8.85&8.13&5.92\\
Colorado&7&11.37&9.87&9.39&9.94&9.5&8.25\\
Connecticut&3&11.71&11.13&11.72&10.34&8.43&6.54\\
District Of Columbia&1&9.19&8.74&7.43&8.25&8.46&7.26\\
Florida&7&5.65&4.23&4.73&5.55&5.34&4.77\\
Georgia&2&11.39&12.15&10.83&11.62&10.8&10.83\\
Hawaii&1&3.04&2.99&4.59&3.21&3.72&2.85\\
Indiana&4&11.49&9.43&8.32&9.89&8.2&7.2\\
Iowa&1&1.42&2.33&1.66&2.08&1.9&1.69\\
Kansas&4&6.10&4.86&4.4&5.98&5.16&4.47\\
Kentucky&1&13.94&14.95&13.08&13.18&17.43&11.25\\
Maine&2&4.30&3.67&3.99&4.03&3.39&2.83\\
Maryland&5&10.43&10&8.57&8.84&8.11&6.74\\
Massachusetts&8&9.47&9.08&7.15&8.45&6.47&5.15\\
Michigan&2&9.37&8.28&8.37&8.59&7.39&6.79\\
Minnesota&2&7.51&5.95&7.25&9.88&6.53&5.08\\
Mississippi&1&3.91&4.25&4.28&3.85&3.79&2.68\\
Missouri&6&10.81&9.17&8.82&8.62&8.64&6.92\\
Montana&1&0.54&0.61&0.9&0.68&0.46&0.36\\
Nevada&2&8.48&7.93&9.77&10.09&8.53&6.76\\
New Jersey&8&13.80&13.39&12.33&12.48&12.38&8.03\\
New Mexico&8&4.68&4.56&4.48&4.74&5.22&3.58\\
New York&5&13.64&12.56&12.05&12.42&11.3&7.55\\
North Carolina&3&6.04&6.29&6.05&6.11&6.58&4.4\\
North Dakota&6&2.91&1.98&2.44&2.59&2.16&1.95\\
Ohio&5&14.60&12.13&10.84&11.17&11.28&8.25\\
Oklahoma&3&8.89&8.43&7.02&7.38&7.38&6.02\\
Oregon&2&12.39&11.59&10.69&9.42&9.98&7.77\\
Pennsylvania&1&12.87&12.89&11.74&10.81&8.7&10.39\\
Rhode Island&2&14.03&13.5&10.23&13.37&10.15&8.47\\
South Carolina&2&5.59&6.03&6.73&6.62&5.84&4.92\\
Texas&40&5.80&6.23&5.23&6.41&5.95&5.26\\
Utah&7&3.42&2.95&4.5&4.15&2.92&2.32\\
Vermont&2&7.32&6.12&6.23&6.16&5.74&4\\
Virginia&9&5.77&5.05&4.92&5.25&5.42&4.04\\
Washington&2&16.80&18.17&14.25&13.32&11.36&9.52\\
Wisconsin&2&12.50&11.11&9.61&11.36&9.75&8.63\\
Wyoming&13&1.69&1.62&1.61&1.55&1.31&1.33\\
   \hline
\end{tabular}
\end{table}

\begin{table}[ht]
\centering
\caption{Monthly-average of \ce{PM_{2.5}} concentrations ($\mu$g/m$^3$) by state in the month of April of the years 2015--2020.}
\label{tab:PM25stats}
\begin{tabular}{l|r|rrrrr|r}
State&N. Sites&2015&2016&2017&2018&2019&2020\\
\hline
Alabama&4&7.94&7.56&8.82&7.26&7.64&8.59\\
Alaska&4&3.45&3.57&4.36&3.43&4.45&3.83\\
Arizona&13&5.67&5.82&6.9&8.13&5.03&4.75\\
Arkansas&4&7.01&7.12&8.12&6.98&7.51&7.54\\
California&61&7.06&6.65&6.35&7.93&6.31&5.26\\
Colorado&8&5.31&3.65&4.63&5.87&5.26&5.21\\
Connecticut&8&4.8&5.54&3.81&5.61&5.32&5.75\\
Delaware&3&5.73&5.65&6.21&6.48&5.91&7.09\\
District Of Columbia&1&6.21&6.68&7.11&8&6.33&4.67\\
Florida&16&7.01&7&7.41&7.68&6.39&9.52\\
Georgia&10&7.87&8.31&8.52&8.25&8.83&8.44\\
Hawaii&7&5.96&4.72&6.68&4.26&3.2&3.32\\
Idaho&5&4.47&5.51&4.21&4.43&3.55&5.72\\
Illinois&14&8.23&7.67&7.06&8.03&7.43&8.11\\
Indiana&15&7.5&8.21&6.17&6.64&6.21&8.47\\
Iowa&9&7.78&6.97&6.18&7.09&5.65&8.71\\
Kansas&3&6.34&6.21&6.85&8.31&8.77&11.47\\
Kentucky&12&6.97&7.03&6.27&6.48&6.77&8.04\\
Louisiana&4&7.81&7.06&8.43&7.52&6.84&7.74\\
Maine&6&5.16&5.64&4.93&4.46&3.63&4.05\\
Maryland&10&6.93&6.75&5.74&6.46&4.33&4.81\\
Massachusetts&9&4.67&5.08&3.02&5.6&4.68&5.71\\
Michigan&11&6.7&6.7&5.52&6.48&6.33&7.1\\
Minnesota&18&5.09&5.66&5.06&6.22&4.88&5.73\\
Mississippi&7&7.98&7.49&8.32&8.27&6.97&10.15\\
Missouri&13&7.65&6.3&6.46&7.33&7.33&6.6\\
Montana&11&5.02&4.57&4.07&4.81&3.7&4.21\\
Nebraska&2&8.02&6.65&9.82&9.77&6.1&8.17\\
Nevada&6&6.16&4.81&4.57&5.25&3.23&4.11\\
New Hampshire&5&4.3&4.2&3.18&4.11&3.34&4.01\\
New Jersey&3&5.69&7.27&7.32&7.17&6.09&5.84\\
New Mexico&5&7.34&5.05&7.02&7.93&5.58&4.87\\
New York&7&5.32&4.99&4.38&4.93&4.92&4.5\\
North Carolina&13&6.78&7.25&7.16&6.53&6.4&5.41\\
North Dakota&6&4.65&3.24&4.33&5.26&3.58&4.2\\
Ohio&12&7.49&7.86&5.55&6.82&6.68&6.99\\
Oklahoma&9&7.32&7.31&7.52&8.78&7.71&8.19\\
Oregon&12&5.3&4.94&4.28&5.14&4.24&5.17\\
Pennsylvania&24&7.41&7.01&7.64&6.99&6.52&6.97\\
Rhode Island&5&4.84&5.2&4.56&6.01&3.3&4.08\\
South Carolina&6&7.33&6.95&7.95&6.64&6.6&6.62\\
South Dakota&8&6.65&4.93&5.35&5.54&3.68&5.01\\
Tennessee&11&6.41&6.68&6.8&6.45&6.5&6.23\\
Texas&11&10.04&8.84&9.13&8.79&8.76&10.27\\
Utah&7&5.6&3.3&3.47&4.59&3.03&3.79\\
Vermont&4&4.32&4.24&3.27&4.69&4.48&4.73\\
Virginia&6&5.74&5.76&6.33&5.31&5.75&5.66\\
Washington&11&5.57&6.5&3.53&4.01&4.17&5.11\\
Wisconsin&18&5.45&7.47&4.58&5.46&6.68&7.49\\
Wyoming&3&4.04&2.75&2.73&3.3&2.57&1.73\\
   \hline
\end{tabular}
\end{table}

\begin{table}[ht]
\centering
\caption{Average air quality measurements in April of the years 2015--2020 from NASA OMI and ground monitoring stations.}
\label{tab:stats}
\begin{tabular}{l|rrrrr|r}
 & 2015 & 2016 & 2017 & 2018 & 2019 & 2020 \\ 
\hline
\textbf{NASA OMI }&&&&&&\\
\ce{NO2} at all pixels (10$^{16}$molecules/cm$^2$) & 0.16 & 0.15 & 0.14 & 0.13 & 0.14 & 0.12 \\ 
\ce{NO2} at ground sites (10$^{16}$molecules/cm$^2$) & 0.32 & 0.31 & 0.28 & 0.29 & 0.29 & 0.23 \\ 
\textbf{Ground monitoring stations}&&&&&& \\ 
\ce{NO2} (ppb) & 8.16 & 7.69 & 7.22 & 7.62 & 7.0 & 5.52\\
\ce{PM_{2.5}} ($\mu$g/m$^3$) & 6.52 & 6.36 & 6.02 & 6.60 & 5.82 & 6.31\\
   \hline
\end{tabular}

\end{table}

\end{document}